 \documentclass[final,3p,times,twocolumn]{elsarticle}
\usepackage{epsfig}
\usepackage{epstopdf}

\usepackage{amssymb}
\usepackage[nodots]{numcompress}

\journal{}

\begin{document}

\begin{frontmatter}



\title{Coherent forward scattering of starlight  by a cloud of atomic hydrogen.}


\author[f1]{Fr\'ed\'eric Zagury} 
\author[p1]{Pierre Pellat-Finet}

\address[f1]{Harvard University,  Cambridge, MA 02138, USA\fnref{f1}}
\address[p1] {LMBA, UMR CNRS 6205, Universit\'e de Bretagne Sud, B.P. 92116, 56321 Lorient cedex, France}
 
\begin{abstract}
Theory predicts that a plane wave scattered by a thin slab of gas yields, in the forward direction and under specific circumstances, a larger irradiance than would be observed in the absence of the gas. 
This enhanced Rayleigh scattering depends on the size of the Fresnel zones at the slab location, as seen from the observer's position, and results from the coherence of the scattering.
On astronomical scales the exceptional size of  Fresnel zones ($\phi\sim1500$~km)  has particular relevance when considering  forward-scattered starlight by an interstellar cloud of atomic hydrogen.
\end{abstract}


\end{frontmatter}

% \linenumbers



%% References

\section{Introduction}
%%%%%%%%%%%%%%%%%%%%%%%%%%%%%%%%%%%%%%%%%%%%%
We consider a light wave of wavelength $\lambda$  that propagates through interstellar space. 
A thin interstellar cloud of atomic hydrogen is interposed on the wave's path.
How will the cloud modify the irradiance of the light an observer measures in the direction of propagation?

In standard astrophysical conditions and at UV wavelengths, theory suggests  (Sects.~\ref{vdh} to \ref{an}) the counter-intuitive result that the irradiance would be much larger than what it would be without the cloud.
Rather than reducing the irradiance at the position of the observer the gas enhances it.
It is this result, the reasons why it does not violate energy conservation and why distances do not appear in the analytical expression of the scattered starlight that we wish to address.

The paper will  be organized as follows.
In Sect.~\ref{vdh} we present an original formula H.C. van de Hulst derived for the scattering of a plane-wave by identical, spherically symmetric, particles in his now classic book on scattering \cite{vdh}.
Before we realized the formulae were identical we had independently reach a similar expression in the more general case of a star at large but finite distance.
We used an alternative method, based on the Huygens-Fresnel theory, which is presented in Sect.~\ref{hf}.
The quantitative comparison of the direct and scattered irradiances is made in Sect.~\ref{an}.
The discussion that follows is based on the relationship which exists between the irradiances from a Huygens-Fresnel sphere and from its first Fresnel zone  (Sects.~\ref{comp}-\ref{sc1}).
This correspondence is in our opinion fundamental.
It helps to explain the  reason why scattering in the forward direction may be enhanced and  provides a different and simpler view on the problem.
Any attempt to derive a more exact expression of the irradiance due to a thin slab of gas  should start with an estimate of the light scattered from the first Fresnel zone alone.
We last investigate conditions for enhanced  scattering in the forward direction and comment on energy conservation  (Sects.~\ref{dis}-\ref{ener}).

Our interest in forward coherent scattering by a gas was motivated by questions pertaining to interstellar extinction. 
Our analysis therefore focuses on the case-study of an interstellar cloud illuminated by a star, although the absence of distances in the expression of the scattered light irradiance may suggest relevant laboratory experiments. 
%%%%%%%%%%%%%%%%%%%%%%%%%%%%%%%%%%%%%%%%%%%%%%%%%%%%%  
\section{Scattering by a cloud of hydrogen from a star at infinity} \label{vdh}
\begin{figure}[t]
\resizebox{\columnwidth }{!}{\includegraphics{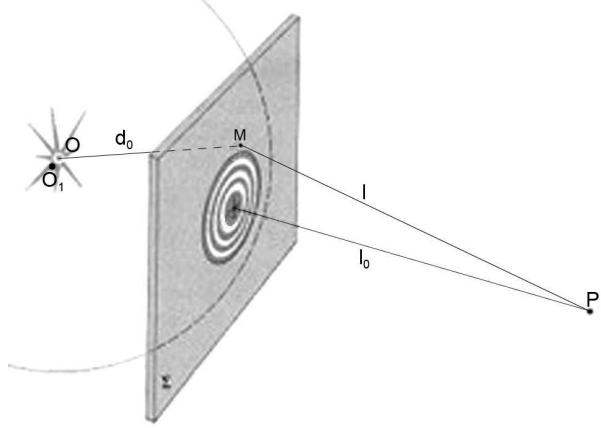}} 
\caption{Schematic representation of the interstellar cloud, the Huygens-Fresnel sphere, and the Fresnel zones (from fig.~10.52 in \cite{hecht}). 
The radius of the Huygens-Fresnel sphere is $d_0$, $d$ is the distance from the source $O$ to a point $M$ in the slab, on or close to the sphere. 
Distances $l$ and $l_0$ are from the observer $P$ to $M$, and to the sphere.
The distance from $P$ to $O$ is $D=l_0+d_0$.
When the star is at infinity (Sect.~\ref{vdh}) the sphere and the plane coincide.} 
\label{fig:fig}
\end{figure}

%%%%%%%%%%%%%%%%%%%%%%%%%%%%%%%%%%%%%%%%%%%%%
A light  wave of wavelength $\lambda$ ($k=2\pi/\lambda$) falls on a cloud of hydrogen atoms with average column density $N_H$ (atom/cm$^2$) represented as a slab $\Sigma$ in Fig.~\ref{fig:fig}.
Let $u_0$  be the amplitude of the disturbance due to the plane wave an observer P would measure in absence of the cloud, and $u$ the amplitude he would measure in the direction of the wave with the cloud on the line of sight.
From sect.~4.3 of van de Hulst's book  'Light scattering by small particles' \cite{vdh}
%%%%%%%%%%%%%%%%%%%%%%%%%%%%%%
\begin{equation}
u=u_0\left(  1-\frac{2\pi}{k^2}N_HS(0)\right),     \label{eq:vdh0}
\end{equation}
%%%%%%%%%%%%%%%%%%%%%%%%%%%%%%%%%%%%%%%%
where $S(0)$ is defined in sect.~6.12 of the same book\footnote{In sect.~6.13 of the same book van de Hulst considers an additional term, $(2/3)k^6\alpha^2$, in the expression of $S(0)$, to account for the diminution of the plane wave because of the scattering. This term only diminishes the direct light from the source and is moreover totally negligible, of order $10^{-8}u_0$ (with the orders of magnitude given in Sect.~\ref{an}). 
It is therefore neglected.}: $S(0)=ik^3\alpha_H$ ($\alpha_H=6.7\,10^{-25}\rm cm^3$ is the polarizability of hydrogen).
Therefore
%%%%%%%%%%%%%%%%%%%%%%%%%%%%%%
\begin{equation}
u=u_0\left(  1-{2\pi}ik \alpha_H N_H\right) \label{eq:vdh1}
\end{equation}
%%%%%%%%%%%%%%%%%%%%%%%%%%%%%%%%%%%%%%%%
The amplitude  $u$  consists of two terms, the direct, unattenuated (provided that $N_H\sigma_\lambda\ll 1$, $\sigma_\lambda=8/3 \pi k^4 \alpha_H^2$ the Rayleigh cross-section of hydrogen at wavelength $\lambda$) light from the source-wave, and a scattered light term, $u_s=-{2\pi}ik \alpha_H N_Hu_0$, which is a quarter of a period out of phase with the primary wave.

Van de Hulst specifies that Eq.~\ref{eq:vdh1} holds only in the  direction of the source-wave, and for a large cloud-observer distance.
He emphasizes that the scattered field at P is primarily influenced by the light scattered by the first Fresnel zones\footnote{The n$\rm^{th}$ Fresnel zone is defined as the set of points M on the Huygens-Fresnel sphere of Fig.~\ref{fig:fig} for which $(n-1)\lambda/2\leq OMP-OP<n\lambda/2$. The first zone is a disc, the following are rings \cite{born}.} as viewed from $P$ (Fig.~\ref{fig:fig}).

%%%%%%%%%%%%%%%%%%%%%%%%%%%%%%%%%%%%%%%%%%%%%%%%%%%%%%%%%%%%%%%%%%%%%
%%%%%%%%%%%%%%%%%%%%%%%%%%%%%%%%%%%%%%%%%%%%%%%%%%%%%%%%%%%%%%%%%%%%%%%%%%%%%%%%%%%%%%%
 %%%%%%%%%%%%%%%%%%%%%%%%%%%%%%%%%%%%%%%%%%%%%%%%%%%%%  
\section{Cloud illuminated by a star at large but finite distance} \label{hf}
%%%%%%%%%%%%%%%%%%%%%%%%%%%%%%%%%%%%%%%%%%%%%
In this section, provided that distances be large enough, we show that the cloud can be assimilated to a Huygens-Fresnel sphere and use Fresnel's theory to re-derive Eq.~\ref{eq:vdh1} in a more general case.
For a negligible thickness $e$ of the cloud ($e\ll l_0$ and $e\ll d_0$) the cloud is a plane which can be approximated  by a portion of the Huygens-Fresnel sphere centered on the star (Fig.~\ref{fig:fig}).
A small area $\mathrm{d}S$ of  $\Sigma$, centered on $M$, contains a large number of scatterers\footnote{Each scatterer is equivalent to a small area $\sigma_\lambda\sim 10^{-24}$~cm$^2$ for hydrogen atoms and at 1500~\AA. In a typical interstellar cloud  $\Sigma$ will have a surface density of $\sim 3.\,10^{20}$  scatterers per cm$^2$ (Sect.~\ref{an}). The disturbance at the observer position due to $\mathrm{d}S$ is $N_H a_0\mathrm{d}S$, with $a_0$ the disturbance due to one atom.} and approximates very well the secondary  sources used in the derivation of Fresnel's theory. 

Huygens considered light as a spherical wave-front perturbation which propagates through space (if the source is at infinity the wave-front is a plane). 
Fresnel found that the contribution of an elementary surface  $ \mathrm{d}S$ centered on $M$ of the wavefront, at distance $d_0$ from the source of light $O$ and $l$ from the observer $P$ (Fig.~\ref{fig:fig}), to the perturbation at $P$ is\footnote{Demonstrations are given in textbooks: \cite[chap.3]{vdh};  \cite[chap.10.3]{hecht};  \cite[chap.8]{born}. The sign of the $e^{i\pi /2}$ factor may change from one textbook to another.}
%%%%%%%%%%%%%%%%%%%%%%%%%%%%%%
\begin{equation}
\mathrm{d}u_f=  \frac{e^{-ikl}}{l\lambda}    \mathrm{d}S \frac{u_0 e^{-ikd_0}}{d_0}e^{i\pi /2}     \label{eq:fresnel}
\end{equation}
%%%%%%%%%%%%%%%%%%%%%%%%%%%%%%%%%%%%%%%%
Only the area  close to axis $OP$ contributes significantly and the obliquity factor is here, as in Fresnel's theory,  neglected. 

On the other hand, if $O$ is a star (Fig.~\ref{fig:fig}), assumed to be point-like, a small area  $ \mathrm{d}S$  close to axis $OP$ of an HI (atomic hydrogen) cloud scatters the star's light and contributes to the disturbance at $P$ by the amount (deduced from \cite[sects.~4.1 and 6.12]{vdh})
%%%%%%%%%%%%%%%%%%%%%%%%%%%%%%
\begin{eqnarray}
\mathrm{d}u_s&=&    k^2 \alpha_H   \frac{e^{-ikl}}{l}   N_H   \mathrm{d}S \frac{u_0 e^{-ikd}}{d}   \nonumber \\
&=& \frac{e^{-ikl}}{l\lambda}    \mathrm{d}S \frac{[4\pi^2u_0 \alpha_H N_He^{-i\pi/2}/\lambda ]e^{-ikd}}{d}e^{i\pi /2}
     \label{eq:scafr}
\end{eqnarray}
%%%%%%%%%%%%%%%%%%%%%%%%%%%%%%%%%%%%%%%%
Amplitude $u_0$ is the perturbation due to the star with no cloud on the line of sight.

If distances are large the sphere and the cloud coincide over a large number of Fresnel zones (Sect.~\ref{cond}).
Identification of the terms in Eq.~\ref{eq:fresnel} with those of Eq.~\ref{eq:scafr}  shows that the cloud is virtually equivalent to a second star, $O_1$, superposed to $O$ and a quarter of a period out of phase with it, with amplitude
 %%%%%%%%%%%%%%%%%%%%%%%%%%%%%%
\begin{equation}
u_1=(4\pi^2\alpha_H N_He^{-i\pi/2}/\lambda) u_0  \label{eq:e2}
\end{equation}
%%%%%%%%%%%%%%%%%%%%%%%%%%%%%%%%%%%%%%%%
The disturbance measured at $P$ is the sum of the disturbances due to $O$ and to $O_1$
 %%%%%%%%%%%%%%%%%%%%%%%%%%%%%%
\begin{equation}
u=u_0\left(  1-{2\pi}ik \alpha_H N_H\right) \label{eq:sca}
\end{equation}
%%%%%%%%%%%%%%%%%%%%%%%%%%%%%%%%%%%%%%%%
This  is for a star at large but finite distance  the same expression as found by van de Hulst (Eq.~\ref{eq:vdh1}).
Van de Hulst's situation is the particular case where the distance to the star  is infinite ($D=l_0+d_0\sim d_0$) and
the Huygens-Fresnel sphere of Fig.~\ref{fig:fig} becomes the slab $\Sigma$.

A real star, if it cannot be considered as a point source, will be split  into small independent sources.
The same result will be reached by adding the contributions of all the sources, provided that $N_H$ remains constant over a sufficiently large area of the cloud.
%%%%%%%%%%%%%%%%%%%%%%%%%%%%%%%%%%%%%%%%%%%%%%%%%%%%%  
\section{Scattered to direct light intensities ratio } \label{an}
%%%%%%%%%%%%%%%%%%%%%%%%%%%%%%%%%%%%%%%%%%%%%%%%%%%%%  
The ratio of the scattered  to the direct irradiances received at $P$ from the direction of the star, if Eq.~\ref{eq:sca}  applies, is
%%%%%%%%%%%%%%%%%%%%%%%%%%%%%%
\begin{equation}
\frac{I_{s}}{I_0}= \left( \frac{|u_s|}{|u_0|}\right)^2= \left(  \frac{ 4\pi^2\alpha_H N_H}{\lambda}\right)^2  \label{eq:r2}
\end{equation}
%%%%%%%%%%%%%%%%%%%%%%%%%%%%%%%%%%%%%%%%
For an interstellar cloud with column density\footnote{The reddening is $\sim0.5$~mag.. The density of these clouds is generally low, a few tens to a few hundreds  atom/cm$^3$.} $N_H=3\,10^{20}\rm cm^{-2}$, at UV wavelength $\lambda$=1500~\AA\
%%%%%%%%%%%%%%%%%%%%%%%%%%%%%%
\begin{eqnarray}
\left|\frac{u_s}{u_0}\right| &  \approx &100  \label{usu0}\\
\frac{I_s}{I_0}&\approx & 10^4
     \label{eq:an}
\end{eqnarray}
%%%%%%%%%%%%%%%%%%%%%%%%%%%%%%%%%%%%%%%%
With these values $N_H\sigma_\lambda\approx 3.5\,10^{-5}\lll1$.
%%%%%%%%%%%%%%%%%%%%%%%%%%%%%%%%%%%%%%%%%%%%%%%%%%%%%  
\section{Discussion} \label{dis}
%%%%%%%%%%%%%%%%%%%%%%%%%%%%%%%%%%%%%%%%%%%%%
%%%%%%%%%%%%%%%%%%%%%%%%%%%%%%%%%%%%%%%%%%%%%%%%%%%%%  
\subsection{Role and importance of the first Fresnel zone} \label{comp}
%%%%%%%%%%%%%%%%%%%%%%%%%%%%%%%%%%%%%%%%%%%%%
Eq.~\ref{eq:sca} does not depend on distances $l_0$ and $d_0$ as long as they are large.
This remarkable fact results from Huygens-Fresnel theory and from the correspondence introduced in Sec.~\ref{hf} between a Huygens-Fresnel sphere and the cloud:
the disturbance  generated by a source of light does not depend on the specific  Huygens-Fresnel sphere chosen between the source and the observer.

Two additional important properties of light propagation are highlighted by Fresnel's theory \cite{hecht, born}. 
First, only the lower order Fresnel zones contribute efficiently to the disturbance.
Second, the contribution of the first zone alone to the disturbance at $P$ is in absolute value twice the overall disturbance itself;
the irradiance at $P$ due to  the first zone, if it can be separated from the contribution of the other zones, is four times the total irradiance\footnote{Poisson thought this effect would  refute Fresnel's memoir \cite{fresnel}. Observations verified the theory and led to the recognition  of Fresnel's work.}.

These properties  imply that for Eq.~\ref{eq:sca} to hold it is enough for the cloud to match an Huygens sphere over a few Fresnel zones only.
They also mean that the scattered irradiance may be estimated from the scattering by the first Fresnel zone alone, and that adding the disturbances from a larger number of  zones diminishes but does not destroy  the contribution of the first zone (unless the cloud matches exactly an even number of Fresnel's zones).

%%%%%%%%%%%%%%%%%%%%%%%%%%%%%%%%%%%%%%%%%%%%%%%%%%%%%  
\subsection{Scattered irradiance from the first Fresnel zone alone} \label{sc1}
%%%%%%%%%%%%%%%%%%%%%%%%%%%%%%%%%%%%%%%%%%%%%
If $d_0$ and $l_0$ are large the irradiance due to scattering by the whole cloud should be one fourth the irradiance  $I_{1}$ in the idealized situation of a cloud confined to the first Fresnel zone.
The surface of a Fresnel zone is \cite{hecht}
 %%%%%%%%%%%%%%%%%%%%%%%%%%%%%%
\begin{equation}
S_{\lambda}=\pi\lambda \frac{d_0l_0}{D} \label{eq:s}
\end{equation}
%%%%%%%%%%%%%%%%%%%%%%%%%%%%%%%%%%%%%%%%
For a given $l_0$, $S_{\lambda}$ is  maximum for a source at infinity ($d_0\approx D$).

Assuming  scattering is isotropic the irradiance $i_{0}$ due to the light scattered by a single atom is
 %%%%%%%%%%%%%%%%%%%%%%%%%%%%%%
\begin{equation}
i_0=\frac{1}{4\pi l_0^2}I_0\left(\frac{D}{d_0}\right)^2\sigma_\lambda = \frac{1}{4\pi}I_0\left(\frac{D}{d_0l_0}\right)^2\sigma_\lambda
 \label{eq:i0}
\end{equation}
%%%%%%%%%%%%%%%%%%%%%%%%%%%%%%%%%%%%%%%%
The number of atoms in the first Fresnel zone is $n_0=N_HS_{\lambda}$.
The order of magnitude $I_1$ of the irradiance due to atoms in the first Fresnel zone alone can be estimated by $I_1=n_0^2i_0$.
From Eqs.~\ref{eq:s} and \ref{eq:i0} and the expression of $\sigma_\lambda$ (Sect.~\ref{vdh})
 %%%%%%%%%%%%%%%%%%%%%%%%%%%%%%
\begin{equation}
\frac{I_{1}}{I_0}=N_H^2S_{\lambda}^2\frac{i_0}{I_0}
=\frac{\pi}{4}\lambda^2\sigma_\lambda N_H^2 
=\frac{2}{3}\pi^2\left(  \frac{ 4\pi^2\alpha_H N_H}{\lambda}\right)^2         \label{eq:i1}
\end{equation}
%%%%%%%%%%%%%%%%%%%%%%%%%%%%%%%%%%%%%%%%
Distances have cancelled as expected. 
Eq.~\ref{eq:i1} and Eq.~\ref{eq:r2} differ by a factor 6.6, close to the factor of 4 anticipated previously.
The difference should be attributed  to the fact that  the phase lags between atoms in the first Fresnel zone have  been neglected.

%%%%%%%%%%%%%%%%%%%%%%%%%%%%%%%%%%%%%%%%%%%%%%%%%%%%%  
\subsection{The effect of distances} \label{dis}
%%%%%%%%%%%%%%%%%%%%%%%%%%%%%%%%%%%%%%%%%%%%%
For a cloud 100~pc~$=3.\,10^{20}$~cm away and a star at infinity (for instance in another galaxy), the first Fresnel zone for $\lambda=1500$~\AA\ is $\sim1500$~km large. 
The amplification of the scattered light  in the forward direction, with respect to what it would be if coherence of the scattered waves was ignored, is then a factor $n_0=5.\,10^{34}$ for $N_H=3.\,10^{20}\rm cm^{-2}$.
If the star is at equal distance from the cloud as the observer is ($l_0=d_0=D/2$), $S_\lambda$ is divided by two, and the irradiance $I_1$ due to the scattered starlight is reduced by a factor of 4 only.
It nevertheless remains very large.

In the expression $I_1\approx n_0^2i_0$  the $n_0^2$ term contributes most to the scattered light when $S_\lambda$ is large  (the cloud is far from both the star and the observer) because of the large number of scatterers and their cooperative effect; $i_{0}$ is then minimum.
Conversely  $i_{0}$ will be enhanced when the scatterers are close to the star or to the observer (because of the $1/(l_0^2d_0^2)$ dependence of $i_0$, Eq.~\ref{eq:i0}), while the size of the Fresnel zone, and thereby coherent scattering, is reduced.

Coherent scattering and scattering by one particle have opposite effects which compensate for each other.
But for stars close to a cloud the curvature of the Huygens-Fresnel sphere is increased and the identification of the sphere with the cloud will be more difficult; in addition the coherence of the scattered waves within the cloud thickness will tend to disappear.
The effect of coherent scattering will be lost and Eq.~\ref{eq:sca} can no more be applied.
In this case the  scattered light  irradiance  should be calculated using classical incoherent scattering.
It is negligible compared to the star irradiance $I_0$.

Only coherent scattering from large Fresnel zones can lead to an appreciable amount of scattered starlight in the sense discussed in the previous sections.
For a given observer-cloud distance this will happen for stars sufficiently far-away from the cloud.
%%%%%%%%%%%%%%%%%%%%%%%%%%%%%%%%%%%%%%%%%%%%%%%%%%%%%  
\subsection{The cloud thickness parameter} \label{cond}
%%%%%%%%%%%%%%%%%%%%%%%%%%%%%%%%%%%%%%%%%%%%%
The difference $\Delta=OMP-D$ between the paths light traverses from $O$ to $P$ via a point $M$  in the cloud  at distance $h$ from  axis $OP$, and $D$ is
%%%%%%%%%%%%%%%%%%%%%%%%%%%%%%
\begin{equation}
\Delta= \frac{h^2D}{2l_0d_0}  \label{eq:delta},
\end{equation}
%%%%%%%%%%%%%%%%%%%%%%%%%%%%%%%%%%%%%%%% 
$\Delta=\Delta_n=n\lambda/2$ gives  the radius $h_n$ of the $n^{\rm th}$ Fresnel zone
%%%%%%%%%%%%%%%%%%%%%%%%%%%%%%
\begin{equation}
\frac{h_n^2 D}{l_0d_0}= n\lambda  \label{eq:fn}
\end{equation}
%%%%%%%%%%%%%%%%%%%%%%%%%%%%%%%%%%%%%%%%
Eq.~\ref{eq:delta} can also be used to find the path-length difference between two points $M_1$ and $M_2$ of the cloud both at distance $h$ from axis $OP $.
Let $(l_0,d_0)$ and  $(l_0+\epsilon,d_0-\epsilon)$ be the distances of the projection of  $M_1$ and $M_2$ on the axis, to $P$ and $O$.
With $l_0\ll d_0$  and $\epsilon\ll l_0$,
%%%%%%%%%%%%%%%%%%%%%%%%%%%%%%
\begin{eqnarray}
\Delta_{1,2}&=&|\Delta_{M_1}-\Delta_{M_2}| \nonumber\\
&=&\frac{h^2D}{2}\left|\frac{1}{l_0d_0}-\frac{1}{\left( l_0+\epsilon \right) \left( d_0-\epsilon \right)}     \right| \\
&\approx&\frac{h^2D}{2d_0l_0^2}\epsilon
    \label{eq:deltad}
\end{eqnarray}
%%%%%%%%%%%%%%%%%%%%%%%%%%%%%%%%%%%%%%%%
If $h\approx h_n$ 
%%%%%%%%%%%%%%%%%%%%%%%%%%%%%%
\begin{equation}
\Delta_{1,2}\approx    \frac{\lambda}{2} \frac{n\epsilon}{l_0}         \label{eq:deltadn}
\end{equation}
%%%%%%%%%%%%%%%%%%%%%%%%%%%%%%%%%%%%%%%%
The scattered waves remain coherent over the thickness $e$ of the cloud as long as  $\Delta_{1,2}$ is less than half the wavelength, that is over $n\approx l_0/e$ Fresnel zones.
%%%%%%%%%%%%%%%%%%%%%%%%%%%%%%%%%%%%%%%%%%%%%%%%%%%%%  
\subsection{Energy conservation} \label{ener}
%%%%%%%%%%%%%%%%%%%%%%%%%%%%%%%%%%%%%%%%%%%%%
The irradiance at $P$ of light scattered by the whole cloud has the same order of magnitude as if the cloud was localized in the first Fresnel zone.
The order of magnitude is that of Eq.~\ref{eq:i1} and must not violate energy conservation:
the power extinguished (by Rayleigh scattering) within the first Fresnel zone must remain much larger than the power that is measured at the focus.

Consider  a small circular surface of area $s_d$ (for instance the surface of a detector) and radius $r_d$ centered at the focus, small enough for the scattered irradiance to vary little across the area.
The power crossing $s_d$  is
 %%%%%%%%%%%%%%%%%%%%%%%%%%%%%%
\begin{equation}
P_{s_d}= s_dI_1=\frac{\pi}{4}\lambda^2\sigma_\lambda N_H^2 s_d I_0      \label{eq:p0}
\end{equation}
%%%%%%%%%%%%%%%%%%%%%%%%%%%%%%%%%%%%%%%%
The power extinguished by $n_0=N_HS_\lambda$ atoms within the cloud (limited to the first Fresnel zone) is
 %%%%%%%%%%%%%%%%%%%%%%%%%%%%%%
\begin{equation}
P_{ext}= N_HS_\lambda\sigma_\lambda I_0      \label{eq:pabs}
\end{equation}
%%%%%%%%%%%%%%%%%%%%%%%%%%%%%%%%%%%%%%%%
Since $P_{ext}\ggg P_{s_d}$
 %%%%%%%%%%%%%%%%%%%%%%%%%%%%%%
\begin{equation}
\frac{4l_0}{N_H\lambda} \ggg s_d      \label{eq:cd}
\end{equation}
%%%%%%%%%%%%%%%%%%%%%%%%%%%%%%%%%%%%%%%%
The area at the focus over which the irradiance of the scattered light may be important (compared to the incoming plane wave irradiance)  will necessarily  be negligibly small in a laboratory experiment.
In astronomical conditions, the inequality of Eq.~\ref{eq:cd} is less restrictive since, with the values of Sect.~\ref{an} and a distance $l_0\sim100$~pc, energy conservation imposes that $ r_d\lll 30$~m ($s_d\lll 1$~km$^2$).

For an infinite slab, as considered by van de Hulst and which is best illustrated by the idealized\footnote{In addition to the gas real interstellar clouds contain a small amount of dust particles which are extremely efficient at extinguishing starlight and introduce a $1/\lambda$ extinction. The attenuation of both the scattered and the direct lights from the star due to these dust particles is neglected here.} interstellar cloud of atomic hydrogen illuminated by a star at large distance, the same calculation holds for any observer at the same distance from the slab.
This is physically justified by the fact that the power extinguished within the slab tends to  infinity.

We also tried to investigate possible  limitations of van de Hulst's formula (Eq.~\ref{eq:vdh1}). 
In the introduction of his Chapter 4 van de Hulst indicates, with no further justification, that the derivations made along the chapter hold as long as the average distance between the particles remains large  compared to the wavelength.
We did not find the reason why it should be so, why classical scattering theory wouldn't apply in standard laboratory conditions, why if $n_0$ atoms were to be localized in the first Fresnel zone the irradiance of the scattered light would not be given by $n^2$ times the irradiance due to one atom?
If   Eqs.~\ref{eq:vdh1}  and \ref{eq:i1} are not applicable to this problem then it remains an open question as to how to calculate the irradiance of the scattered light.
In interstellar space however average densities are extremely low (Sect.~\ref{an}) and van de Hulst's formula is fully justified.
%%%%%%%%%%%%%%%%%%%%%%%%%%%%%%%%%%%%%%%%%%%%%%%%%%%%%%  
\section{Conclusion } \label{conc}
%%%%%%%%%%%%%%%%%%%%%%%%%%%%%%%%%%%%%%%%%%%%%
This paper is based on a formula (Eq.~\ref{eq:vdh1}) which provides the irradiance of a plane wave scattered by a slab of identical, spherically symmetric, particles, in the forward direction and at the position of an observer far-away from the slab.
The formula in itself presents no special difficulty.
It may be obtained by different methods, through direct integration over all the particles in the slab (as it was first derived  by van de Hulst in 1957), or using the convenient and more visual framework of the Huygens-Fresnel theory.
These methods  use no more than straightforward principles of optics and general scattering theory. 
The Huygens-Fresnel construction highlights specific aspects of Eq.~\ref{eq:vdh1}, its non-dependence on distances, and the role of the first Fresnel zone.
Fresnel's theory also relates the irradiance at the position of the observer due to the whole slab (Eq.~\ref{eq:r2}) and the irradiance due to solely the first Fresnel zone (approximated by Eq.~\ref{eq:i1}).
Both lead to the same order of magnitude for the irradiance of the scattered light.

Numerical application of  Eq.~\ref{eq:r2} or Eq.~\ref{eq:i1} in the case of Rayleigh scattering by hydrogen atoms has not been carried out before.
It leads to a surprisingly  large and counter-intuitive ratio of scattered to direct light  irradiances.
The symmetry of atoms, the size of the first Fresnel zone, the coherence of the scattering from atoms in the first Fresnel zone, are the determinant factors which contribute to this large ratio.
In order to be compared with real measurements theory may need to be refined and its conditions of application evaluated in a more precise way than we did.  
But it does suggest the possibility for the image of a star to appear brighter when observed behind a cloud of hydrogen than it would be without the cloud on the line of sight.
The cloud, rather than acting as a screen, behaves as a lens and enhances the irradiance of the light coming from the direction of the star.

The scattering of starlight by an interstellar cloud was the major focus of this paper.
We have shown that this problem  is equivalent to the coherent scattering from the first Fresnel zone. 
If $n_0$ atoms are enclosed in the first Fresnel zone, the irradiance at the focus should roughly be the product of $n_{0}^2$ by the irradiance $i_0$ one atom alone would give: two atoms will give $4i_0$ at the focus (four times the irradiance of one atom), three  $9i_0$, and so forth.
A discussion of Eq.~\ref{eq:vdh1} (van de Hulst's formula) can therefore be simplified by considering first the idealized situation of a cloud localized to the first Fresnel zone.
Reciprocally this particular case  will be used to understand how  theory needs to be improved before it can be compared with observation.
How will distances between the source, the slab, and the observer, or the thickness of the cloud, intervene? 
Do, as suggested by van de Hulst, distances between atoms need to be considered, and if so how will  Eq.~\ref{eq:i1} and Eq.~\ref{eq:vdh1} be modified? 

Our goal was to call attention on a specific case of scattering which received little consideration.
The underlying physics is extremely simple but the consequences seem to have passed unnoticed.
We have outlined that Fresnel's zones can be unusually large on astronomical scales but the absence of distances in Eq.~\ref{eq:vdh1} may allow laboratory experiments which would provide an insight into the questions forward coherent scattering by a gas  can raise. 
%%%%%%%%%%%%%%%%%%%%%%%%%%%%%%%%%%%%%%%%%%%%%%%%%%
\section*{Acknowledgments}
%%%%%%%%%%%%%%%%%%%%%%%%%%%%%%%%%%%%%%%%%%%%%
Fr\'ed\'eric Zagury was supported by an Arthur Sachs Fellowship, and is grateful for  
the hospitality and resources provided by Harvard University.

\bibliographystyle{model3-num-names}
{}
\end{document}